\documentclass[12pt]{article}
\usepackage{amsmath}
\usepackage{amssymb}
\usepackage{graphicx}
\usepackage{epsfig}
\usepackage[all,knot,poly,dvips]{xy}

\begin{document}

\begin{center}
{ \large \bf
BF--theory in graphene:\\
a route toward topological quantum computing? }
\end{center}

\vspace{12pt}

\begin{center}
{\large {\em A. Marzuoli}$^{\,\dag, \, \S,\,(1)}$ and {\em G. Palumbo}$^{\, \ddag, \,\S,\,(2)}$}
\end{center}

\vspace{6pt}

\begin{center}
{\small $\dag\,$ Dipartimento di Matematica `F. Casorati',
  Universit\`a degli Studi di Pavia,\\
via Ferrata 1, 27100 Pavia (Italy)\\
$\ddag$ Dipartimento di Fisica Nucleare e Teorica,
Universit\`a degli Studi di Pavia,\\
via A. Bassi 6, 27100 Pavia (Italy)\\
$\S\,$ Istituto Nazionale di Fisica Nucleare, Sezione di Pavia,\\
via A. Bassi 6, 27100 Pavia (Italy)\\
$^{\,(1)}$ E-mail: annalisa.marzuoli@pv.infn.it \\
$^{\,(2)}$ E-mail: giandomenico.palumbo@pv.infn.it
}
\end{center}

\vspace{6pt}

\begin{abstract}
Besides the plenty of applications of graphene allotropes in
condensed matter and nanotechnology, we argue that graphene sheets
might be engineered to support  room--temperature topological
quantum processing of information. The argument is based on the
possibility of modelling the monolayer graphene effective action
by means of a 3d Topological Quantum Field Theory of BF--type able
to sustain non--Abelian anyon dynamics. This feature is the basic
requirement of recently proposed theoretical frameworks for
fault--tolerant and decoherence protected quantum computation.
\end{abstract}

\vspace{2 cm}

\vfill
\newpage

\section{Introduction}

The issue of connections between topological quantum field
theories (TQFT) in three spacetime dimensions \cite{Witt2,Ati}
and condensed matter systems in $d =$ $2,3$ has been
intensively investigated over the years in a variety of different
contexts. The occurrence of topological phases of matter in ground
states and the existence of quasi--particle excitations
associated with
fractionary statistics \cite{Wilc} are the crucial features of
many--body microscopic systems sharing topological effective
actions. Such  collective behaviors have been observed in
the Fractional Quantum Hall Effect, cold atoms in optical
lattices, Topological Insulators and graphene sheets
\cite{Bern,Ser,Pach}. Most recent proposals of
experimental settings have also fostered theoretical research
aimed to classify classes of models supporting topological phases
\cite{Lev,Kit}.

Topological (or anyonic) quantum computing is a promising
territory where models and tools from topological (effective)
quantum field theory might find new exciting applications
\cite{Freed1,Freed3}. The basic ingredients of any anyonic--type
computation involve: i) the choice of a finite set particle types,
i.e. labels specify the possible values of the {\em charges}; ii)
the assumption that particles can fuse and split according to a
set of rules that give the charge of a composite particle in term
of the constituents; iii) the assumption that particles
trajectories are braided according to rules specifying how pairs
(or bipartite subsystems) behave under exchange. 3d TQFT of the
Schwarz--type \cite{Bir,Kaul}, as well as 2d (boundary) Conformal
Field Theories and charge sectors of families of
(2+1)--dimensional gauge theories (with finite or compact Lie
gauge group) are theoretical frameworks able to support
multi--dimensional unitary representation of the braid group
obeying fractionary statistics. However, the requirements of
fault–-tolerance and stability under local quantum perturbations
--achieved when every unitary gate can be approximated within any
precision by braiding and fusing transformations-- selects more
restrictive classes of theories. For instance \cite{Freed2}, {\em
doubled} non-Abelian Chern-Simons (CS) theories supporting
non-Abelian statistics, provide  such a PT--invariant theoretical
framework. This circumstance provides the basic motivation of this
letter since any 3d doubled Chern--Simons TQFT can be converted
into a BF--theory, a fact that has been already exploited in a
number of recent approaches to infrared regimes of many--body
quantum systems \cite{Moor,Blas,Ard,Fen,Han}.

The goal of this work is to exploit the doubled CS
$\leftrightarrow$ BF correspondence in order to provide
theoretical evidence for the emergence of non--Abelian anyons in
monolayer graphene, thus opening the challenging opportunity for a
`room temperature' topological quantum computer. In the present
context non--Abelian anyons are induced by some amount of disorder
due to random impurities and by resorting to the replica method it
is shown that the effective action of the system is given by a
non--Abelian BF--type action
with $U(2)$ playing the role of the gauge group (for any choice of
the level $\kappa>0$). The two emergent basic fields $\mathbf{A}$
and $\mathbf{B}$, the gauge potential (connection) and the
so--called $B$--field to be associated with the presence of
vortices, turn out to be related in a natural way to their
microscopic counterparts.
Finally, the issue of (gauge invariant) quantum observables to be
associated with the effective theory is briefly addressed.

It is
worth stressing that, unlike pure CS, the BF framework –-originally
formulated in terms of continuous geometric structures and as such
used in the following-– possesses natural discretized counterparts
given by the class of the Turaev-–Viro state sum models ({\em cfr.} a few
more remarks in the last section). The issue of giving an
equivalent lattice formulation of the effective BF action for
graphene is crucial for assessing and improving the potentialities
of the model in view of realistic applications for quantum
computational purposes. Progress in this direction is under study
and will be presented elsewhere.


\section{Graphene:  microscopic description}

Graphene \cite{Novo} is a two--dimensional planar honeycomb lattice with
strongly bounded carbon atoms placed at the sites, and the physical properties
of this material are under intense scrutiny \cite{Been,Neto}.

 At room temperature, near the Fermi points, the charge carries exhibit a
 relativistic behavior and thus their wave functions are suitably described
 by a (2+1)--dimensional massless Dirac equation \cite{Seme}. This  property
 is related to the fact that a vanishing gap between the
 conductance and valence bands in the energy spectrum has been experimentally
measured.\\
The massless Dirac equation is explicitly given by
\begin{equation}\label{dirac1}
  i \, \gamma^{\mu} \; \partial_{\mu} \; \psi = 0
\end{equation}
where $\mu=0,1,2$
and the bi--spinor is
\begin{equation}\label{dirac2}
  \psi=\left(
\begin{array}{c}
\psi_{A}^{+} \\
\psi_{B}^{+} \\
\psi_{A}^{-} \\
\psi_{B}^{-}
\end{array}
\right),
\end{equation}
where $A$ and $B$ are labels denoting the two triangular sublattices
of the honeycomb lattice and $+$ and
$-$ refer to the two distinct Fermi points. Both  types of
indices represent internal
degrees of freedom of the charge carriers in graphene. Consequently
the $\gamma^{\mu}$ are $4\times4$ Dirac matrices and in the following
the chiral representation will be always used.

It is possible to
 get a gap in several different manners (see for instance \cite{Ryu}).
 Here we are going to introduce a chemical
 potential $\mu$ into the Dirac equation.

The next ingredient of the construction is to
take into account  vortices generated by topological
defects in the honeycomb lattice \cite{Guine,Jack,Fran,Chamo}.
 To this end, we introduce
$U(1) \times U(1)$ gauge fields $a_{\mu}$ and $b_{\mu}$
coupled with fermions, to be
identified with the ordinary electromagnetic potential and the chiral
gauge field, respectively \cite{Chamo,Ryu}.
Thus the  action of the resulting coupled system reads
\begin{equation}\label{action1}
 S [\psi, \overline{\psi}; a_{\mu},b_{\mu} ]\,=\,-\kappa \int d^{3}x \; \overline{\psi}_{s}\,
( i \gamma^{\mu} \partial_{\mu} -\gamma^{\mu}
 \; a_{\mu}-\gamma_{5}\, \gamma^{\mu}b_{\mu} \,-\, \mu\gamma_{0})\, \psi^{s}
\end{equation}
where $\kappa$ is for the moment an arbitrary constant and we have
included a further index $s=1,2$ which takes into account the
(real) spin degeneracy of particles \cite{Son}. It is worth noting
that  the form of  this action resembles  the one considered  in
\cite{Snyde} in connection with Topological Insulators. The final
ingredient amounts to include properly the presence of
(unavoidable) disorder on the graphene sheet  \cite{Neto2,Ostro}.
Following the argument of the authors of \cite{Snyde}, we do not
add an explicit disorder potential but rather exploit the replica
method, a tool employed already in graphene monolayer, see
\cite{Alt,Zieg}. This means  that we can introduce $N$ replicas in
the action ~(\ref{action1}) where  fermions are coupled with
$U(N)\times U(N)$ gauge fields denoted by
\begin{equation}\label{fields}
\mathbf{a}_{\mu}=a_{\mu}^{\alpha}T_{\alpha};\;\;\;
\mathbf{b}_{\mu}=b_{\mu}^{\alpha}T_{\alpha},
\end{equation}
where $T_{\alpha}$ are the (Hermitian)
generators of $U(N)$.


\section{Euclidean effective action}

The Euclidean (Wick rotated)  counterpart of the action
~(\ref{action1}) written in terms of the fields (\ref{fields})
is obtained by performing the following
substitutions
\begin{equation}\label{eucl}
  t\rightarrow i\tau;\hspace{1.0cm}
\gamma^{\mu}\partial_{\mu}\rightarrow i\gamma^{\mu}\partial_{\mu};\hspace{1.0cm}
  \gamma^{\mu}\mathbf{a}_{\mu}\rightarrow i\gamma^{\mu}\mathbf{a}_{\mu}\hspace{1.0cm}
\gamma^{\mu}\mathbf{b}_{\mu}\rightarrow i\gamma^{\mu}\mathbf{b}_{\mu}.
\end{equation}

By integrating out the fermionic fields, the Euclidean
effective action depends on the two gauge fields alone
\begin{equation}\label{eff0}
\int
\mathcal{D}[\psi_{1}, \overline{\psi}_{1},\psi_{2}, \overline{\psi}_{2}]\;
e^{-S}
\,=\,e^{-S_{\text{eff}}\,[\mathbf{a}_{\mu},\mathbf{b}_{\mu}]}.
\end{equation}

The derivative expansion of $S_{\text{eff}}$ reads
\begin{equation}\label{eff1}
S_{\text{eff}} \,=\,-Tr \ln (G_{0}^{-1})+\sum_{n=1}^{\infty}\frac{1}{n}Tr
[G_{0}(-i\gamma^{\mu}\mathbf{a}_{\mu}-i\gamma^{\mu}\mathbf{b}_{\mu})]^{n},
\end{equation}
where $G_{0}$ is the propagator of free Dirac fermions.

At leading order and upon applying a
Pauli--Villars regularization, we obtain
\begin{equation}\label{eff2}
\tilde{S}_{\text{eff}\,}\,=\,\kappa
\left(\frac{2}{4\pi}I[\mathbf{A}^{+}]-\frac{2}{4\pi}I[\mathbf{A}^{-}]\right)
\end{equation}
where the $\mathbf{A}_{\mu}^{\pm}$ (in terms of anti--Hermitian generators) are related
to the original gauge fields by
\begin{equation}\label{eff3}
\mathbf{A}_{\mu}^{\pm}=\mathbf{a}_{\mu}\pm \mathbf{b}_{\mu} \,,
\end{equation}
and where (tr is the trace over Lie algebra labels and $\epsilon^{\mu\nu\sigma}$
is the Levi--Civita totally antisymmetric symbol)
\begin{equation}\label{eff4}
I[\mathbf{A}_{\mu}^{\pm}]= \int d^{3}x \; \epsilon^{\mu\nu\sigma} \,\text{tr}\,
\left(\mathbf{A}_{\mu}^{\pm}\partial_{\nu}\mathbf{A}_{\sigma}^{\pm}+
\frac{2}{3} \mathbf{A}_{\mu}^{\pm}
\mathbf{A}_{\nu}^{\pm}\mathbf{A}_{\sigma}^{\pm}\right).
\end{equation}

Looking at the numerical constant inside the bracket in
~(\ref{eff2}) and comparing it with the standard form of
Chern--Simons action, $\tilde{S}_{\text{eff}}$ turns out to be
proportional (trough $\kappa$) to a double non--Abelian
$U(N)_{2}\times \overline{U(N)}_{2}$ CS at level $k=2$ (recall
that the level represents the quantized CS coupling constant, an
integer multiple of $(4\pi)^{-1}$, and the overline means that the
second Chern--Simons term has opposite chirality).

By resorting to results found in the 90's about the level--rank
duality \cite{Nacu,Nacu2}, considered here in a CS environment,
the exchange between the rank $N$ of the gauge group and the level
$k$ provides a consistent dualized $U(2)_{N}\times
\overline{U(2)}_{N}$ action. A dual model shares the same fusion
rules, modular transformation matrices and observables of its
parent theory, so that the two are to be considered as equivalent
also in the present context of effective field theories, as
pointed out in \cite{Freed2}(Section III).

On applying to the partition function $Z$
associated with ~(\ref{eff2}) the formal $N\rightarrow 0$ limit we get
\begin{eqnarray}\label{eff5}
\overline{S_{\text{eff}}}:=\frac{d}{d N}Z^{N}|_{N=0}=\kappa
\frac{d}{d N} \frac{2}{4 \pi}
\left\{I[\mathbf{A}^{+}]-I[\mathbf{A}^{-}]\right\}|_{N=0}=&& \nonumber \\
\kappa \frac{d}{d N} \frac{N}{4 \pi}
\left\{I[\mathbf{A}^{+}]-I[\mathbf{A}^{-}]\right\}|_{N=0}=\frac{\kappa}{4
\pi}\left\{I[\mathbf{A}^{+}]-I[\mathbf{A}^{-}]\right\},
\end{eqnarray}
where the switch $N\leftrightarrow k$ has been explicitly included.

The so far unrestricted effective coupling constant $\kappa$  is
now required to assume only integer values according to the
standard argument used whenever the Feynman quantization
prescription is going to be carried out in a topological field
theory background. Thus ~(\ref{eff5}) represents a
$U(2)_{\kappa}\times \overline{U(2)}_{\kappa}$  double CS action
endowed with an `effective level' $\kappa$. This topological field
theory has been shown to be equivalent (both at the classical and
at the quantum level) to a BF--type  theory
\cite{Bir,Marte,Arch,Witt} with a classical action given by
\begin{equation}\label{BF1}
S_{BF,\,\lambda}\,=\,\int d^{3}x \; \, \epsilon^{\mu\nu\sigma} \, \text{tr}
\left(\mathbf{B}_{\mu}\mathbf{F}_{\nu\sigma}+
\frac{\lambda ^2}{3} \;\mathbf{B}_{\mu}\mathbf{B}_{\nu}\mathbf{B}_{\sigma}\right).
\end{equation}
Here
$\lambda$ is a constant, related to $\kappa$ by
\begin{equation}\label{lambda}
\lambda^2=\left(\frac{4
\pi}{\kappa}\right)^{2},
\end{equation}
$\mathbf{F}_{\nu\sigma}$ is the curvature
2-form associated with the connection 1-form $\mathbf{A}_{\mu}$  according to
\begin{equation}\label{effe}
\mathbf{F}_{\nu\sigma} \,=\, \partial_{\nu}\mathbf{A}_{\sigma}-
\partial_{\sigma}\mathbf{A}_{\nu}+[\mathbf{A}_{\nu},\mathbf{A}_{\sigma}]  ,
\end{equation}
$\mathbf{B}_{\mu}$ (the B--field) is canonically conjugate to $\mathbf{A}_{\mu}$ and
\begin{equation}\label{BF2}
  \mathbf{A}_{\mu}=\frac{1}{2}(\mathbf{A}^{+}_{\mu}+\mathbf{A}^{-}_{\mu})
 \hspace{1.0cm}
  \mathbf{B}_{\mu}=\frac{\kappa}{8\pi}(\mathbf{A}^{+}_{\mu}-\mathbf{A}^{-}_{\mu}).
\end{equation}

The standard terminology `BF action with a (positive) cosmological constant term $\lambda^2 \,$'
for the action  ~(\ref{BF1}) bears on the fact that, once chosen a 3-dimensional, ~
compact and oriented Riemannian manifold
$M^3$, up to a suitable isomorphism  this action would be mapped into the (first--order form of)
Euclidean Einstein--Hilbert action of General Relativity (given by the first integral)
plus a term proportional to the volume of the underlying spacetime $M^3$.
The relevant  gauge group would be  $SO(3)$ (or its universal covering $SU(2)$) and
the fields would have a  geometric nature: the field strength $\mathbf{F}_{\nu\sigma}$
can be related to (a contraction of) the Riemann tensor and the B--field
to the dreibein (an orthonormal set of three basis vectors expressed in suitable
local coordinates).


\section{Observables}

Coming back to the interpretation of ~(\ref{BF1})
as an effective action for graphene, the original gauge fields
 are related through ~(\ref{eff3}) to the BF fields by
\begin{equation}\label{BF3}
\mathbf{a}_{\mu}  =
  \mathbf{A}_{\mu}
\hspace{1.0cm}
  \mathbf{b}_{\mu} = \frac{4\pi}{\kappa} \mathbf{B}_{\mu}\,\equiv\, \lambda \mathbf{B}_{\mu},
\end{equation}
so that the BF framework complies with the description
at the microscopic level. Thus the $\mathbf{A}$--field
is consistently interpreted as the (non--Abelian counterpart of the)
electromagnetic potential, while the `chiral' $\mathbf{B}$--field
bears on the presence of vortices. Geometrically the
$\mathbf{A}$--field is still a connection,
while the 1-form $\mathbf{B}$ is related to
1-dimensional submanifolds embedded into the 3-dimensional background.
(Note that  the shift in front of  $\mathbf{B}_{\mu}$ in
~(\ref{BF3}) does not alter the nature of this field, a fact that
would not be true for a connection 1-form).\\

Quantum observables in a BF framework are found on applying the
standard machinery of 3d TQFT. First, the generating functional is formally given
by the path integral
\begin{equation}\label{ZBF}
Z_{BF,\, \lambda} [M^3]\,=\,
\int \mathcal{D}\mathbf{A}
\mathcal{D}\mathbf{B}
\exp \, \{i\, S_{BF,\,\lambda}\, (\mathbf{A}, \mathbf{B})\}
\end{equation}
for a fixed background 3-manifold $M^3$ (the integration domain
in the expression of $S_{BF,\,\lambda}$ in ~(\ref{BF1})).
In a purely geometric, field theoretic context $M^3$ is a closed,
oriented Riemannian manifold and it can be shown that the functional
~(\ref{ZBF}) is a topological invariant of the manifold,
related to the square of the modulus of
the Witten--Chern--Simons \cite{Witt,Tur,Marte}. In the present context
of 2d graphene sheets embedded into a (2+1)--dimensional background
suitable boundary (edge) terms should be taken into account
explicitly.

Gauge invariant quantum observables are associated with embedded oriented (closed)
curves $C \subset M^3$. More precisely \cite{Marte}, vacuum expectation
values of observables in an $SU(N)$ BF are given formally by
\begin{equation}\label{Obser}
Z_{BF,\, \lambda} [M^3,\, C]\,=\,
\int \mathcal{D}\mathbf{A}
\mathcal{D}\mathbf{B}
\exp \,\{i\, S_{BF,\,\lambda}\, (\mathbf{A}, \mathbf{B})\}
\, \text{tr}\, \text{Hol} (\mathbf{A} \pm \lambda \,\mathbf{B}),
\end{equation}
where tr is over Lie algebra labels and
Hol $(\mathbf{A} $ $\pm \lambda \,\mathbf{B})$
are holonomies
\begin{equation}\label{Hol}
 \text{Hol} (\mathbf{A} \pm \lambda \,\mathbf{B})
 \,=\, \exp \, \left\{i \int_C \, (\mathbf{A}_{\mu} \pm \lambda \,\mathbf{B}_{\mu})\,
dz^{\mu} \right\}
\end{equation}
evaluated (up to path ordering)
along the curve $C$ parametrized by local coordinated $z^{\mu}$.

We argue that the quantization of the $U(2)_{\kappa}$ BF
setting described in the previous section will provide explicit expressions for
quasi--particles excitations associated with closed paths surrounding
vortices. Issues that should be worked out include the choice of
boundary conditions, the selection of a proper gauge fixing
and possibly  perturbative expansions of observables
in terms of powers of  $(4 \pi/\kappa) = \lambda$ for $\lambda \rightarrow 0$.
Work is in progress to improve these developments.


\section{Concluding remarks}

The detection of topological effects in graphene represents, on
the one hand, a major challenge for graphene physics (possibly
also in view of applications to topological quantum computing) and
an ideal playground for testing geometrical models and methods
\cite{Pach}, on  the other. Looking in particular at the
morphology of graphene, there have been observed  non--planar
arrangements of bent graphene sheets, carbon nanotubes, fullerenes
and also schwarzites \cite{Ben} which are respectively associated
with cylindrical, spherical and hyperbolic configurations. Carbon
nanocones \cite{Heib} are associated with singular ({\em i.e.} not
smooth) surfaces.
The BF setting presented in this paper seems particularly suitable
to model also the effective behavior of samples of graphene sheets
with different intrinsic geometry and equipped with a variety of
boundary conditions.

As mentioned in the introduction, the crucial
feature that makes BF-theory so promising in the quantum
computational context is given by the equivalence of BF quantum
functionals with Turaev--Viro (TV) state sum models \cite{Tur}.
  The latter provide an {\em ab initio} discretized and colored ambient
  3-manifolds possibly endowed with (piecewise linear) graphs or
loops (the colorings on the 3d triangulation and on 1-dimensional
subsets are induced by elements of the $SU(2)_{\kappa}$
representation ring). Thus the continuous geometric picture
outlined in the previous sections would become fully discretized,
with functional integrals replaced by combinatorial, finite-–type
state functionals to be associated with ground states and
(evolving) edge or point--like configurations carrying fractionary
charges. The TV setting has been already addressed in connection
with abstract  models for topological quantum computation
\cite{Mar1,Mar2,Koe}. A proper inclusion of the effective behavior
of graphene in such an unified scheme would then represent a major
theoretical achievement as well as a viable tool for explicit
evaluations of significant physical quantities.


\section*{Acknowledgments}

{\small We are grateful to Giorgio Benedek, Giancarlo Jug, Zoltan
K\'ad\'ar and Silvano Garnerone for their comments. We are in debt
with the referees for suggestions that have improved the
presentation of the paper.}

\end{document}